\documentclass[a4paper,11pt]{article}
\pdfoutput=1

\usepackage{jheppub}
\usepackage{graphicx}
\usepackage{amsmath}

\newcommand{\eq}[1]{eq.~\eqref{eq:#1}}
\newcommand{\eqs}[2]{eqs.~\eqref{eq:#1} and \eqref{eq:#2}}
\renewcommand{\sec}[1]{section~\ref{sec:#1}}

\newcommand{\fig}[1]{figure~\ref{fig:#1}}

\newcommand{\mycites}[1]{refs.~\cite{#1}}
\newcommand{\mycite}[1]{ref.~\cite{#1}}

\newcommand{\lqcd}{\Lambda_\mathrm{QCD}}

\allowdisplaybreaks[2]

\newcommand{\ord}[1]{{\mathcal O}(#1)}

\newcommand{\ORD}[1]{{\mathcal O}\biggl(#1\biggr)}

\newcommand{\Mae}[3]{\bigl\langle#1\bigr\rvert#2\bigr\rvert#3\bigr\rangle}

\newcommand{\bnslash}{\bar{n}\!\!\!\slash}
\newcommand{\bare}{\mathrm{bare}}

\newcommand{\bn}{\bar{n}}

\newcommand{\df}{\mathrm{d}}

\newcommand{\Li}{\mathrm{Li}}

\newcommand{\eps}{\epsilon}

\newcommand{\cL}{{\mathcal L}}
\newcommand{\cI}{{\mathcal I}}

\newcommand{\nn}{\nonumber}

\newcommand{\hp}{\hat{p}}
\newcommand{\bnP}{\overline {\mathcal P}}

\newcommand{\conv}{\!\otimes\!}
\newcommand{\convz}{\!\otimes_z\!}

\newcommand{\zero}{{(0)}}
\newcommand{\one}{{(1)}}
\newcommand{\two}{{(2)}}

\newcommand{\ktsq}{\vec{k}_{\perp}^{\,2}}

\newcommand{\cusp}{\mathrm{cusp}}

\title{The Fully-Differential Quark Beam Function at NNLO}

\author{Jonathan R.~Gaunt and}
\author{Maximilian Stahlhofen}

\affiliation{Theory Group, Deutsches Elektronen-Synchrotron (DESY), Notkestra\ss e 85
, D-22607 Hamburg, Germany}

\emailAdd{jonathan.gaunt@desy.de}
\emailAdd{maximilian.stahlhofen@desy.de}

\abstract{We present the first calculation of a fully-unintegrated parton distribution (beam function) at next-to-next-to-leading order (NNLO).
We obtain the fully-differential beam function for quark-initiated processes by matching it onto standard parton distribution functions (PDFs) at two loops.
The fully-differential beam function is a universal ingredient in resummed predictions of observables probing both the virtuality as well as the transverse momentum of the incoming quark in addition to its usual longitudinal momentum fraction.
For such double-differential observables our result provides the part of the NNLO singular cross section related to collinear initial-state radiation (ISR), and is important for the resummation of large logarithms through N$^3$LL.
}

\keywords{QCD, NNLO Calculations, Hadronic Colliders, Jets}

\begin{document}
{\flushright DESY 14-170\\
\today\\[-9ex]}
\maketitle

\section{Introduction}
\label{sec:intro}

Fully-differential beam functions (dBFs) are generalized (unintegrated) PDFs that in addition to the Bjorken-variable $x$ depend on the transverse momentum ($\vec k_\perp$) relative to the beam axis and the transverse virtuality ($k^+k^-\!= -t < 0$) of the colliding parton. 

We assume both of these scales to be perturbative, and that the ISR forms a jet roughly along the direction of the incoming beam with the jet axis deviating from the beam axis by a small angle $\sim|\vec k_\perp|/k^-$.
More precisely, we consider the kinematic situation, where $Q^2 \gg t \sim \ktsq \gg \lqcd^2$ and $Q \sim k^-$ is the scale associated with the hard partonic process. This is the regime for which dBFs appear in factorization/resummation formulae for cross sections, see e.g.~\mycite{Jain:2011iu}.\footnote{Other kinematic regimes are possible, e.g. $ Q^2 \gg t^2/Q^2 \sim \ktsq \gg \lqcd^2$. In this region one requires beam functions differential only in $k_\perp$ (TMD PDFs) and a fully differential soft function rather than the dBFs. Reference~\cite{Larkoski:2014tva} discusses the issue of interpolating between these regimes for related double-differential cross sections.} The dBFs are independent of the hard process and depend only on the properties of the colliding parton. They describe the effects of the collinear ISR on (double-)differential cross section measurements probing the full four-momentum $k^\mu$ of the parton that enters the hard interaction. Because the total invariant mass of the ISR jet must be non-negative, $\ktsq$ is constrained for fixed $t$ as~\cite{Jain:2011iu}
\begin{align} \label{eq:ktsqlimits}
\frac{1-x}{x} \, t \,\ge\, \ktsq \,\ge\,0\,. 
\end{align}

The beam functions we are concerned with in this work can be formally defined as proton matrix elements of operators in soft-collinear effective field theory (SCET)~\cite{Bauer:2000ew, Bauer:2000yr, Bauer:2001ct, Bauer:2001yt, Bauer:2002nz, Beneke:2002ph}. 
The first type of beam function to be defined in this way was the virtuality-dependent beam function~\cite{Stewart:2009yx, Stewart:2010qs}.
This beam function was generalized to include transverse momentum dependence in \mycite{Mantry:2009qz}, although in that paper the beam functions are functions of the Fourier-conjugate variable to transverse momentum, i.e. the impact parameter, (and named iBFs).
We will however use the momentum-space definition of the quark dBF given in \mycite{Jain:2011iu}:
\begin{align} \label{eq:dBq_def}
B_q(t, x, \ktsq ) &= \theta(k^-)\Mae{p_n(p^-)}{\bar{\chi}_n(0)\, \delta(t \!- k^-\hp^+)  \frac{\bnslash}{2} 
\bigl[\delta(k^- \!\!- \bnP_n) \frac{1}{\pi}\delta(\ktsq \!- \vec{\mathcal{P}}_{n\perp}^2) \chi_n(0) \bigr]}{p_n(p^-)}.
\end{align}
Here $p_n(p^-)$ denotes the incoming spin-averaged proton state with lightlike momentum $p^\mu=p^- n^\mu/2$, $x\equiv k^-/p^-$ and $\chi_n$
is the gauge-invariant $n$-collinear quark field operator in SCET.
Since we do not measure the polarization of the quark initiating the hard process, the quark dBF does not depend on the orientation of the (two-dimensional) vector $\vec k_\perp$~\cite{Jain:2011iu}. 
We use the usual light-cone (Sudakov) decomposition for four-vectors: $q^\mu = q^- n^\mu/2 + q^+ \bn^\mu/2 + q_\perp^\mu$,
with $n^2 = \bn ^2 = 0$, $n \cdot \bn = 2$.
The delta functions in \eq{dBq_def} measure the label transverse and minus momentum of the quark. The respective SCET label momentum operators are $\vec{\mathcal{P}}_{n\perp}$ and $\bnP_n$~\cite{Bauer:2001ct}.  $\hp^+$ is the plus-momentum operator acting on all fields (including the proton state) to the right.
For more details on the relevant SCET notations and conventions, we refer to \mycites{Stewart:2009yx, Stewart:2010qs, Gaunt:2014xga}.

Particles with momentum $q$ are $n$-collinear if their momentum components scale as $(q^+, q^-, q_\perp) \sim q^-(\lambda^2, 1, \lambda)$, where $\lambda \ll 1$ is the power expansion parameter of SCET. For the calculation of the dBFs $\lambda^2 \sim t/Q^2$.
From the above kinematics it is clear that for fixed $t$ the only propagating degrees of freedom that can interact with the incoming parton carry either $n$-collinear momenta $\sim Q(\lambda^2, 1, \lambda)$  or (ultra)soft momenta $\sim Q(\lambda^2, \lambda^2, \lambda^2)$. 
The proper effective field theory (EFT) setup in this case is SCET$_{\rm I}$. 
SCET$_{\rm I}$ is also used when only the virtuality $t$ is measured~\cite{Stewart:2009yx, Stewart:2010qs}.
(The measurement of $k^-$, i.e. the $x$-dependence of the beam functions is always understood.)
On the other hand, if the observable is only sensitive to the partonic transverse momentum $k_\perp$, modes with momenta $\sim Q(\lambda, \lambda, \lambda)$ are the relevant soft degrees of freedom and the appropriate EFT framework is SCET$_{\rm II}$. 
Like the virtuality-dependent beam functions, but unlike the beam functions only differential in $k_\perp$ (TMD PDFs), the dBFs therefore do not require an extra rapidity regulator (after zero-bin subtractions~\cite{Manohar:2006nz}). The rapidity regularization for SCET$_{\rm II}$ problems is discussed e.g. in \mycites{Chiu:2009yx,Becher:2011dz,Chiu:2012ir}. 
In our calculation of the quark dBF the divergences of any Feynman diagram, ultraviolet (UV) or infrared (IR), are regulated by dimensional regularization ($d=4-2\eps$) only (and zero-bin contributions vanish as scaleless integrals). 
For a more detailed discussion of this issue and a comparison of the dBFs to similar concepts of unintegrated PDFs in perturbative QCD~\cite{Collins:2007ph,Rogers:2008jk}, we refer to \mycite{Jain:2011iu}. 

As for the less differential beam functions, we can perform an operator product expansion (OPE) for the dBFs in SCET~\cite{Fleming:2006cd, Stewart:2009yx}:
\begin{align} \label{eq:dBi_OPE}
  B_i(t,x,\ktsq,\mu) 
  & = \sum_j \int_x^1 \frac{\df z}{z}
   \cI_{ij}\Big(t,\frac{x}{z},\ktsq,\mu\Big) f_j(z,\mu)
   \bigg[1+ \ORD{\frac{\lqcd^2}{t},\frac{\lqcd^2}{\ktsq}}\bigg]\,.
\end{align}
For $t \sim \ktsq \gg \lqcd^2$ we then obtain the dBFs by computing the matching functions $\cI_{ij}(t,z,\ktsq,\mu)$ perturbatively and convolving them with the standard PDFs $f_j(z,\mu)$.
Integrating this perturbative result for the dBF over the full range of $\ktsq$ given by \eq{ktsqlimits} yields the virtuality-dependent beam function $B_i(t,x,\mu)$:
\begin{equation} \label{eq:dBi_int}
 \int \!\! \df^2k_\perp \, B_q(t,x,\ktsq,\mu) = \pi \!\int \!\! \df \big(\ktsq\big) \, B_q(t,x,\ktsq,\mu) = B_q(t,x, \mu)\,.
\end{equation}
It is however impossible to deduce the perturbative TMD PDF $B_i(x,\ktsq,\mu)$ from a simple integration of the renormalized dBF over $t$~\cite{Jain:2011iu}. 
This is because, unlike the $k_\perp$-integral, the $t$-integral is not constrained by the kinematics, \eq{ktsqlimits}, and diverges indicating that the $t$-integration and the regularization of UV (and rapidity) divergences does not commute. The proper derivation of the TMD PDF requires implementing a rapidity regulator and performing the integration over $t$ in the bare dBF before taking the $d\to4$ limit.

In \mycite{Jain:2011iu}, the full set of the dBF matching coefficients $\cI_{ij}$ was calculated at one loop (correcting an earlier result in \mycite{Mantry:2010mk}).
Moreover it was shown, that the renormalization group (RG) evolution of the dBFs is the same as for the virtuality-dependent beam functions, which in turn equals the one of the (virtuality-dependent) jet function~\cite{Stewart:2010qs}.
Since the noncusp anomalous dimension of the jet function~\cite{Stewart:2010qs,Berger:2010xi} as well as the cusp anomalous dimension~\cite{Korchemsky:1987wg,Moch:2004pa} are known to three loops, the RG running of the dBFs is known through N$^3$LL. 
The only missing piece in the N$^3$LL RG resummation kernel is the four-loop correction to the cusp anomalous dimension, which however can be expected to have an almost negligible numerical impact for  processes at present colliders, see \mycite{Gaunt:2014cfa}.

On top of that, N$^3$LL precision for the full differential cross section also requires the NNLO fixed-order expressions for the relevant beam, soft, hard and jet functions.
It is the aim of the present paper to determine the coefficients $\cI_{ij}$ for the (anti)quark dBF ($i=q,\bar q$) at two loops. 
Besides the two-loop results for the virtuality-dependent beam functions~\cite{Gaunt:2014xga,Gaunt:2014cfa} and TMD PDFs~\cite{Gehrmann:2012ze,Gehrmann:2014yya}
our calculation extends the set of quark beam functions available at NNLO.

Higher order results for the dBFs may eventually help to systematically improve the initial state parton shower of Monte Carlo event generators beyond LL~\cite{Collins:2004vq,Collins:2005uv, Watt:2003mx, Watt:2003vf,Alioli:2012fc,Alioli:2013hqa}.
Other possible applications are precise predictions of transverse momentum distributions in Drell-Yan-like processes with a veto on hard central jets, where the jet veto is achieved by a cut on
a virtuality-sensitive observable~\cite{Tackmann:2012bt}. A related factorization formula is discussed in section 4.1 of~\mycite{Jain:2011iu}.\footnote{We note however that this particular factorization formula, in which the jet veto is achieved by using a cut on the global beam thrust, is incomplete and should receive leading power corrections from Glauber modes \cite{Gaunt:2014ska}. This can be avoided by a local veto on an exclusive jet-algorithm-based observable~\cite{Tackmann:2012bt}, where the effects of Glaubers is $\ord{R^2}$ suppressed, with $R$ the jet radius.}
Last but not least the quark dBF plays a prominent role for exclusive $N$-jet production in DIS. In \mycite{Kang:2013nha}, three different (thrust-like) $1$-jettiness event shape variables $\tau_1^{a,b,c}$ were defined, see also \mycites{Antonelli:1999kx,Kang:2013wca,Kang:2013lga}, and the corresponding factorization formulae were derived.\footnote{Reference~\cite{Antonelli:1999kx} discusses the variable $\tau_1^b$ under the name ``DIS thrust''.} The factorization formulae for $\tau_1^{b,c}$ involve the quark dBF. Our NNLO result for the quark dBF represents the last important ingredient to improve the corresponding resummed two-jet predictions in DIS from NNLL to N$^3$LL precision.

The outline of this paper is as follows. In \sec{match} we sketch our two-loop matching calculation for the quark dBF and point out the main differences to our calculation of the NNLO virtuality-dependent beam functions~\cite{Gaunt:2014xga,Gaunt:2014cfa}. Section~\ref{sec:results} contains the novel results for the dBF matching coefficients and \sec{conclusions} our conclusions.

\section{Calculation}
\label{sec:match}

Our calculation of the two-loop quark dBF matching coefficients closely follows our calculation for the virtuality-dependent quark beam function in~\mycite{Gaunt:2014xga}, see also \mycite{Gaunt:2014cfa} for further details. Due to the large overlap of the calculations, we will restrict ourselves to discussing the important differences rather than going through the whole calculation in detail. Since QCD is charge conjugation invariant the antiquark matching coefficients can easily be obtained from the quark ones according to $\cI_{\bar q q} = \cI_{q \bar q}$ and $\cI_{\bar q g} = \cI_{q g}$ ($q=u,d,s,\ldots$), so we will only consider the quark coefficients in the remainder of this section.

\begin{figure}[t]
\begin{center}
\includegraphics[width=0.245\textwidth]{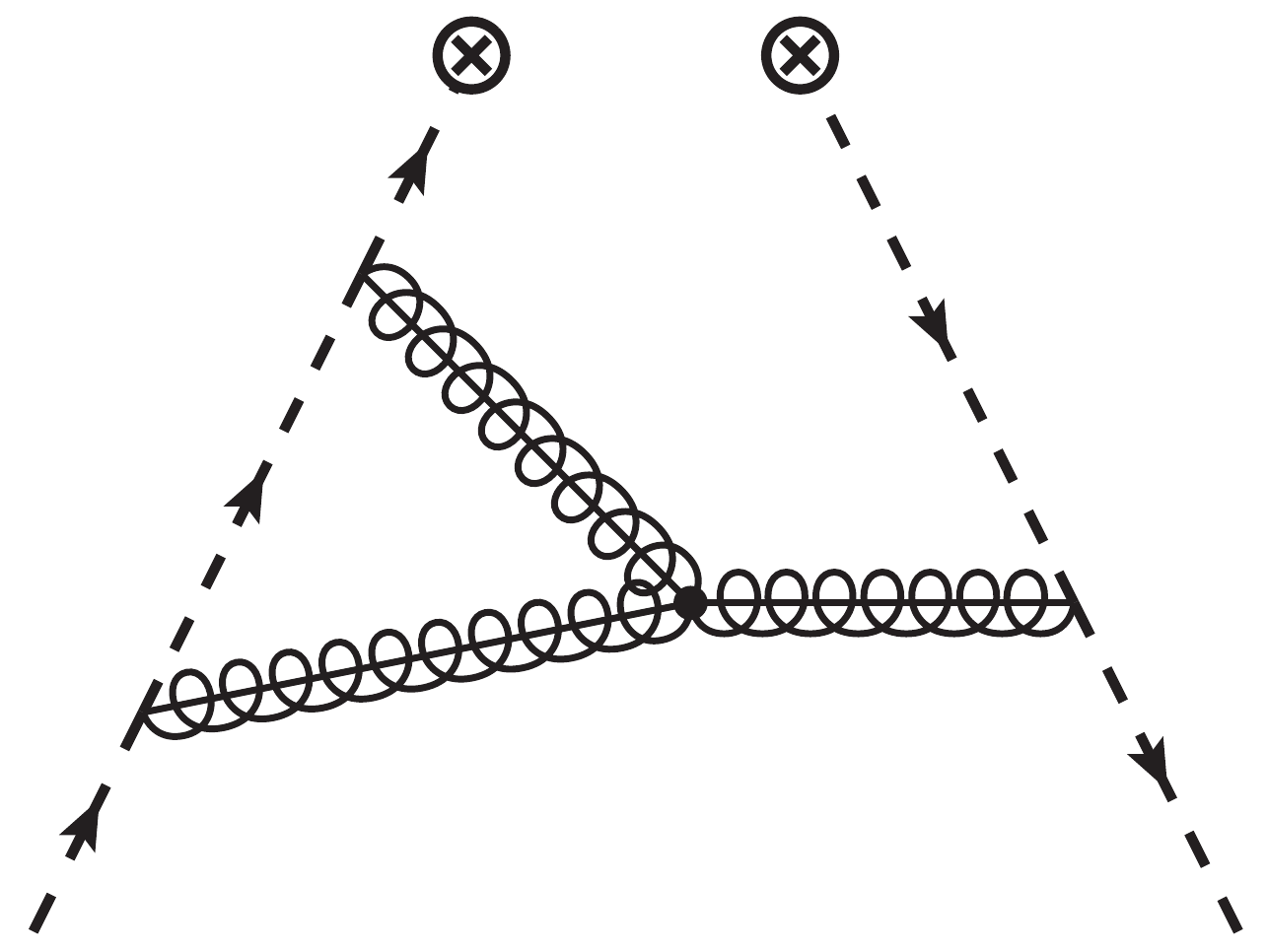}
\qquad
\includegraphics[width=0.245\textwidth]{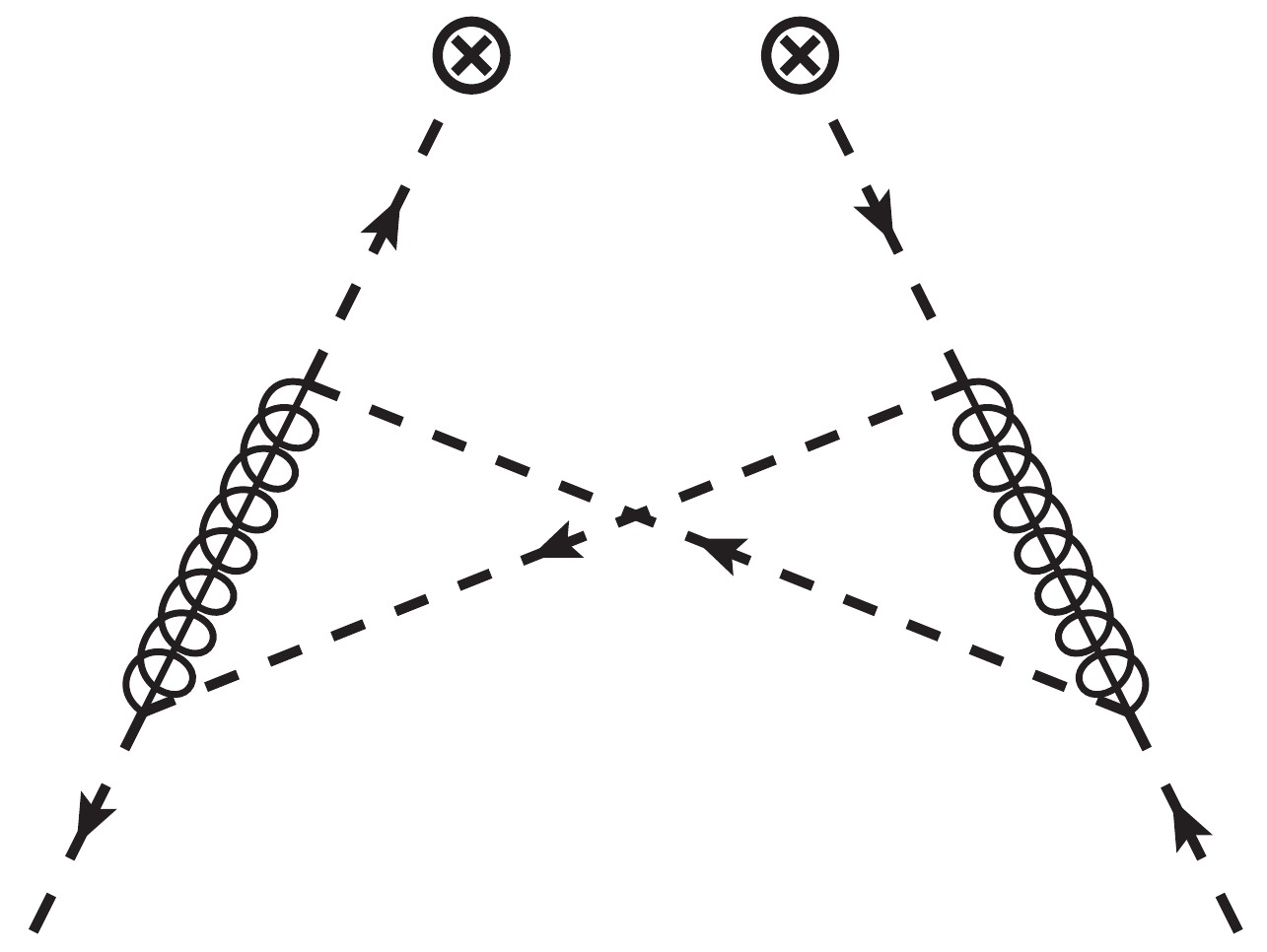}
\qquad
\includegraphics[width=0.245\textwidth]{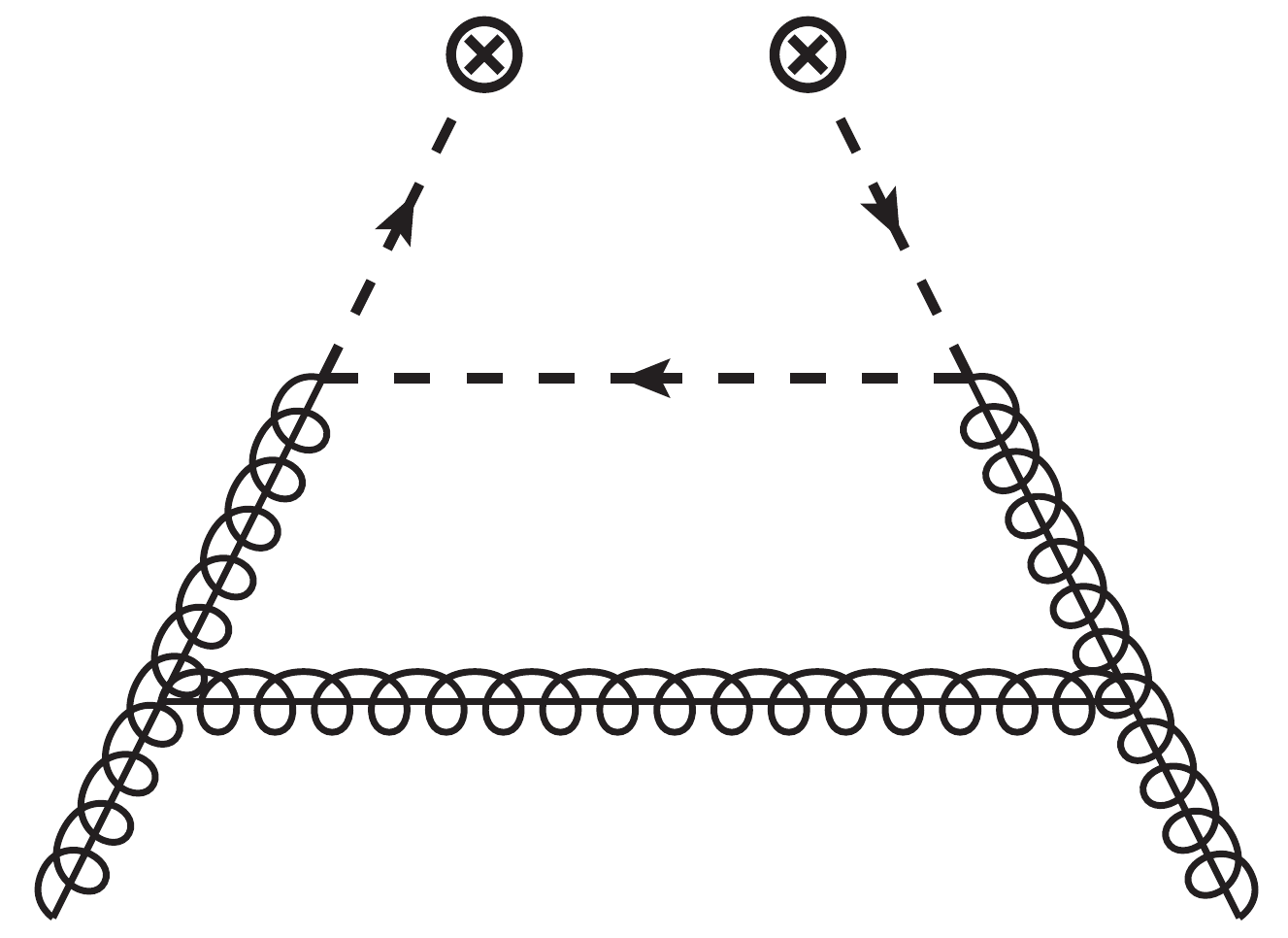}
\end{center}
\caption{Examples of Feynman diagrams contributing to the calculation of the NNLO matching coefficients $\cI_{q_iq_j}$, $\cI_{q_i\bar{q}_j}$ and $\cI_{q_ig}$, respectively. The complete list of relevant axial-gauge diagrams, when using dimensional regularization, is displayed in figure~2 of \mycite{Gaunt:2014xga}.
\label{fig:ExampleDiags}}
\end{figure}

We begin by computing the bare two-loop quark beam function in a partonic state $j$ ($j = q, \bar{q}, g$), which is defined by the matrix element of the same bare dBF operator in \eq{dBq_def}, but with the incoming proton replaced by the parton $j$. We denote this by $B_{q/j}^\bare (t,z, \ktsq)$. In accordance with \mycites{Stewart:2010qs,Gaunt:2014xga,Gaunt:2014cfa} we denote the light-cone minus-momentum fraction by $z$ rather than $x$ when using a partonic state $j$. The kinematic constraint \eq{ktsqlimits} obviously also holds on the partonic level and hence for both $x$ and $z$. 

To obtain the $B_{q/j}^\bare (t,z, \ktsq)$ we calculate the discontinuity of two-loop diagrams like the ones shown in \fig{ExampleDiags}. The complete set of diagrams relevant in axial-gauge and dimensional regularization is given in \mycite{Gaunt:2014xga}, where now the bilocal operator represented by the two $\otimes$ symbols also measures $\ktsq$ according to \eq{dBq_def}. The bare partonic dBF is  related to the renormalized one $B_{q/j}(t,z, \ktsq,\mu)$ by ($i=q$)
\begin{align}  \label{eq:B_ren}
  B_{i/j}^\bare (t,z, \ktsq) & = \int_0^t\! \df t'\, Z_B^i(t - t',\mu)\, B_{i/j}(t',z, \ktsq,\mu)\,.
\end{align}
with $Z_B^i$ the same renormalization factor as for the virtuality-dependent beam function~\cite{Jain:2011iu}. This relation holds on the operator level, i.e. independently from the state $j$ and hence also for the physical proton dBF $B_i(t,z, \ktsq,\mu)$. 

Finally the matching coefficients $\cI_{ij}$ may be extracted using the partonic analog of \eq{dBi_OPE}. The PDFs in the partonic state $j$ are given up to the two-loop order we need in eqs.~(2.22) and~(2.23) of \mycite{Gaunt:2014xga}.
Upon integration over $\vec k_\perp$, both the bare and the renormalized partonic dBFs as well as our final results for the $\cI_{ij}$ must yield the respective results for the virtuality-dependent beam function. At each step this serves us as a strong cross check of the independently obtained parts of the two calculations.

We evaluate the Feynman diagrams together with taking their discontinuity (i.e. performing the unitarity cut) using two methods -- the ``On-Shell Diagram Method'' and the ``Dispersive Method'', which are described in detail in \mycites{Gaunt:2014xga, Gaunt:2014cfa}. We also use two different gauges -- namely light-cone axial ($\bar{n} \cdot A_n = 0$) gauge and Feynman gauge. The two different methods and gauge choices gave the same final results, hence providing us with a strong cross check. As in the previous calculations, the calculation of the bare beam function is done in two stages -- first we compute the beam function away from $z = 1$, $\tilde{B}^{\bare}_{q/j}(t,z,\ktsq)$, and then we add the endpoint $z \to 1$ contribution, $\delta(1-z)D_{q/j}(t,\ktsq)$. 

However, in this calculation we do not need to calculate the endpoint contribution explicitly -- we can extract it from our previously-calculated bare virtuality-dependent beam function as follows. Since the endpoint contribution is proportional to $\delta(1-z)$, it must also be 
proportional to $\delta(\ktsq)$ according to the constraint \eq{ktsqlimits} ($z$ playing the role of $x$), i.e.~$\delta(1-z)D_{q/j}(t,\ktsq) = \delta(1-z)\delta(\ktsq)D_{q/j}(t)$. The integral of the fully differential beam function over $\ktsq$ must give the virtuality-dependent beam function, also at the bare level, leading to
\begin{align}
\pi\,\delta(1-z) D_{q/j}(t) + \int \!\! \df^2k_\perp \, \tilde{B}^{\bare}_{q/j}(t,z,\ktsq) = B^{\bare}_{q/j}(t,z)\,.
\end{align}
All terms in this equation apart from $D_{q/j}(t)$ are known, so we can use this equation to extract $D_{q/j}(t)$ and therefore the endpoint.

Note that in our previous calculation of the bare virtuality-dependent quark beam function, the integrations over the transverse components of the loop momenta and the expansion in the dimensional $\eps$ were performed before the integrations over the loop minus components.
This means that we cannot trivially obtain the dBF by taking our previous calculation and undoing the last integration. However, many of the integral results obtained in that calculation could be re-used in the present context and no additional tool was needed to carry out the integrations for the dBF.

One result required to simplify the piece of our bare partonic dBF proportional to $\delta(t)$ is the following distributional identity (which holds if the function $f$ is integrable):
\begin{align} \label{eq:distId}
\delta(t)\, \theta \Big(t\dfrac{1-z}{z} - \ktsq\Big) \, \dfrac{1}{t}\, f\bigg(\frac{\ktsq}{t}\bigg) 
= \delta(t)\,\delta(\ktsq) \int_0^{\frac{1-z}{z}} \!\!\! \df r f(r) \,.
\end{align}
Note that the $\theta$-function on the left hand side of \eq{distId} is present in all terms of the bare partonic dBF (that are regular in the argument of this $\theta$). It technically originates from the unitarity cut through the two-loop diagrams, similarly to the $\theta(1-z)$ in the virtuality-dependent beam function calculation, and enforces the constraint analogous to \eq{ktsqlimits}. 
Together with the $\delta(t)$ the $\theta$-function restricts the integration range for $\ktsq$ and $\ktsq$ itself to zero on the left hand side of \eq{distId}.
The $t\to0$ limit of the last three factors on the left hand side therefore gives a $\delta(\ktsq)$ normalized by the integral on the right hand side of \eq{distId}.
The correctness of \eq{distId} can be verified by integrating both sides of the equation over $\ktsq$.

The $\delta(t)$ piece of the bare partonic dBF as obtained directly from the two-loop calculation outlined above has the form of the left hand side of \eq{distId}. 
We therefore conclude that it is the same as the $\delta(t)$ piece of the bare partonic virtuality-dependent beam function, up to a factor of $\delta(\ktsq)/\pi$. 
It is actually not surprising that one can use the virtuality-dependent beam function to predict both the $\delta(t)$ and the $\delta(1-z)$ pieces of the dBF, given the similar role of $t$ and $(1-z)$ in the constraint \eq{ktsqlimits} (for $x=z$).
In fact \eq{distId} also holds for $\delta(t)$ replaced by $\delta(1-z)$ on both sides of the equation.

The renormalization scale ($\mu$) dependent terms in the matching functions $\cI_{ij}(t,z,\ktsq, \mu)$ are fixed by solving the corresponding renormalization group equation (RGE),
\begin{align}
\label{eq:IRGE}
\mu\frac{\df}{\df\mu}\cI_{ij}(t,z,\ktsq, \mu)
= \sum_k \int\!\!\df t' \;  & \cI_{ik}(t\!-\!t', z,\ktsq, \mu) 
 \nn \\
& \convz  \Bigl[\gamma_B^i(t', \mu)\,  \delta_{kj} \delta(1\!-\!z) 
 - 2 \delta(t') P_{kj}(z,\mu) \Bigr],    
\end{align}
where the standard Mellin convolution in the minus-momentum fraction is denoted by $\,\convz\,$ and defined in \eq{convz}.
The function
\begin{align}
\gamma_B^i(t, \mu)
&= -2 \Gamma^i_{\cusp}[\alpha_s(\mu)]\,\frac{1}{\mu^2}\cL_0\Bigl(\frac{t}{\mu^2}\Bigr) + \gamma_B^i[\alpha_s(\mu)]\,\delta(t)
\end{align}
is the full (virtuality-dependent) beam function anomalous dimension, and $P_{kj}(z,\mu)$ is the QCD splitting function.
The anomalous dimension $\gamma_B^i(t, \mu)$ equals the jet function anomalous dimension $\gamma_B^i(t, \mu) = \gamma_J^i(t, \mu)$~\cite{Stewart:2010qs}. The various cusp ($\Gamma^i_n$) and non-cusp contributions ($\gamma_{B\,n}^i$) are collected up to three loops ($n=2$) for the quark case ($i=q$) in appendix A.1 of \mycite{Gaunt:2014xga}.
The terms in the perturbative expansion of the splitting function $P^{(n)}_{ij}$ can be found up to NLO ($n=1$) in appendix A.3 of \mycite{Gaunt:2014xga}. (We also list the LO expression ($n=0$) in appendix~\ref{app:splittingfunc}.)

Let us expand the matching coefficient as follows (note the overall $1/\pi$ factor compared to our corresponding expansion for the integrated version of $\cI_{ij}$ in \mycites{Gaunt:2014xga,Gaunt:2014cfa}):
\begin{align}
\cI_{ij}(t,z,\ktsq,\mu) \,=\, \frac1\pi\, \sum_{n=0}^{\infty} \left(\dfrac{\alpha_s}{4\pi}\right)^{\!n} \cI_{ij}^{(n)}(t,z,\ktsq,\mu)
\,.\end{align}
The tree level and one-loop terms, $\cI_{ij}^{(0)}$ and $\cI_{ij}^{(1)}$, are given in the appendix in \eq{Iijtree} and \eq{Iijoneloop}, respectively.

Solving the RGE, \eq{IRGE}, iteratively to NNLO yields the master formula for the two-loop matching coefficient:
\begin{align} \label{eq:I2master}
\cI_{ij}^\two(t,z,\ktsq,\mu)&=
 \frac{1}{\mu^2} \cL_3\Bigl(\frac{t}{\mu^2}\Bigr) \frac{(\Gamma_0^i)^2}{2}\, \delta_{ij}\,\delta(1-z)\, \delta(\ktsq) \nn \\[1 ex] & 
  + \frac{1}{\mu^2} \cL_2\Bigl(\frac{t}{\mu^2}\Bigr)
  \Gamma_0^i \biggl[- \Bigl(\frac{3}{4} \gamma_{B\,0}^i + \frac{\beta_0}{2} \Bigr)\, \delta_{ij}\,\delta(1-z)\, \delta(\ktsq)  \nn\\&\qquad
  + 2P^\zero_{ij}(z)\, \delta(\ktsq) + P^\zero_{ij}(z)\, \delta\Bigl(t\frac{1\!-\!z}{z}-\ktsq\Bigr) \biggr]
  \nn \\[1 ex] &
  + \frac{1}{\mu^2} \cL_1\Bigl(\frac{t}{\mu^2}\Bigr)
  \biggl\{ \Bigl[\Gamma_1^i - (\Gamma_0^i)^2 \frac{\pi^2}{6} + \frac{(\gamma_{B\,0}^i)^2}{4} + \frac{\beta_0}{2} \gamma_{B\,0}^i \Bigr]\,\delta_{ij}\,\delta(1-z)\, \delta(\ktsq)
  \nn \\ & \qquad
  - \Big[\gamma_{B\,0}^i + 2 \Gamma_0^i \ln\!\Big(\frac{1\!-\!z}{z}\Big)\Big] P^\zero_{ij}(z)\,\delta(\ktsq) 
  + 2\Gamma_0^i\, I^\one_{ij}(z)\,\delta(\ktsq) \nn\\&\qquad
  - \Big[\gamma_{B\,0}^i + 2 \beta_0 + 2 \Gamma_0^i \ln\!\Big(\frac{1\!-\!z}{z}\Big)\Big] P^\zero_{ij}(z)\,\delta\Bigl(t\frac{1\!-\!z}{z}-\ktsq\Bigr)\nn\\&\qquad
  + 2\Gamma_0^i \Big[\theta \Big(t\dfrac{1-z}{z} - \ktsq\Big) \frac{1}{t} \cL_0\Bigl(\frac{\ktsq}{t}\Bigr)  + \frac{1}{t} \cL_0\Bigl(\frac{1\!-\!z}{z}-\frac{\ktsq}{t} \Bigr) \Big] P^\zero_{ij}(z) \nn \\[1 ex] & \qquad
  + 4 \sum_k  \Big[\delta\Bigl(t\frac{1\!-\!z}{z}-\ktsq\Bigr) P^\zero_{ik}(z) \Big] \conv_z P^\zero_{kj}(z) \biggr\} \nn \\[1 ex] & 
  + \frac{1}{\mu^2} \cL_0\Bigl(\frac{t}{\mu^2}\Bigr)\, 4 J^\two_{ij}(t,z,\ktsq)
  \,+\, \delta(t)\,\delta(\ktsq)\, 4 I^\two_{ij}(z)\,,
\end{align}
where $\beta_0 = (11 C_A - 4 T_F n_f)/3$, and
\begin{align} \label{eq:plusdefmaintxt}
\cL_n(x)
&= \biggl[ \frac{\theta(x) \ln^n x}{x}\biggr]_+
 = \lim_{\eps \to 0} \frac{\df}{\df x}\biggl[ \theta(x- \eps)\frac{\ln^{n+1} x}{n+1} \biggr]
\end{align}
defines the usual plus distributions.
All ingredients in \eq{I2master} were explained and given for the quark case ($i=q$) in \mycite{Gaunt:2014xga}, except for the functions $J^\two_{ij}(t,z,\ktsq)$ and the $z$-convolutions in the $\cL_1(t/\mu^2)$ term. The results for the latter convolutions are presented for $i=q$ in appendix \ref{sec:oneloopconvs}. The $I^\two_{ij}(z)$ functions here are the same ones appearing in the matching coefficient for the virtuality-dependent beam function. This is because the $\delta(t)$ pieces of $B^{\bare}_{q/j}(t,z,\ktsq)$ and $B^{\bare}_{q/j}(t,z)$ are equal (up to a factor of $\delta(\ktsq)/\pi$) as shown above and the respective renormalization factor $Z_B^q$ is the same. 

The functions $J^\two_{ij}(t,z,\ktsq)$ for $i=q$ are the novel outcome of our dBF calculation. Despite being the coefficient of $\cL_0(t/\mu^2)$ we cannot predict the $J^\two_{ij}$ from the RGE, only their integral over $\ktsq$. The reason for this is that when we differentiate the $\cL_0(t/\mu^2)$ term in \eq{I2master} with respect to $\mu$, we obtain [using \eq{Lnderivs}]:
\begin{align}
\mu \dfrac{d}{d\mu} \frac{1}{\mu^2} \cL_0\Bigl(\frac{t}{\mu^2}\Bigr)\, 4 J^\two_{ij}(t,z,\ktsq) 
&= -8 \delta(t) J^\two_{ij}(t,z,\ktsq) 
\\ \nn
&= -8 \delta(t) \delta(\ktsq) \!\int_0^{\frac{1-z}{z}} \!\df r \, J^\two_{ij}(t,z, t\, r)\,,
\end{align}
where in the second line we use \eq{distId}. We see that by comparing this term to the corresponding $\delta(t)$ term on the right hand side of the RGE, \eq{IRGE}, we can only extract the integral of $J^\two_{ij}(t,z,\ktsq)$ over $\ktsq$. We present results for the $J^\two_{ij}(t,z,\ktsq)$ in the case $i=q$ (and $i = \bar q$) in the next section.

The diagonal components ($i=j=q$) of our master formula, \eq{I2master}, and also the results for the $J^\two_{ij}$ contain terms that might appear ill-defined as $z\to1$ or $t\to0$ at first sight.
To obtain a meaningful result in these limits one is however forced to perform the integration over $r\equiv \ktsq/t$ analogous to \eq{distId} first. This will generate terms that exactly cancel the ones ill-defined in the $z\to1$ or $t\to0$ limits leaving only regular contributions and well-defined distributions.
As an example consider the ominous terms in \eq{I2master},
\begin{align}
\label{eq:exampleterms}
\frac{1}{\mu^2} \cL_1\Bigl(\frac{t}{\mu^2}\Bigr) 2 \Gamma_0^q &\bigg[\! -\ln\!\Big(\frac{1\!-\!z}{z}\Big)\, \delta(\ktsq) -  \ln\!\Big(\frac{1\!-\!z}{z}\Big)\, \delta\Bigl(t\frac{1\!-\!z}{z}-\ktsq\Bigr) \nn\\&
+\theta \Big(t\dfrac{1-z}{z} - \ktsq\Big) \frac{1}{t} \cL_0\Bigl(\frac{\ktsq}{t}\Bigr)  + \frac{1}{t} \cL_0 \Bigl(\frac{1\!-\!z}{z}-\frac{\ktsq}{t} \Bigr) \bigg]P^\zero_{qq}(z)
\end{align}
with $P^\zero_{qq}(z) \propto \cL_0[(1-z)/(1+z^2)]$, see appendix~\ref{app:splittingfunc}.
In the limit $z\to1$ (or $t\to0$) we have
\begin{align}
 \delta\Bigl(t\frac{1\!-\!z}{z}-\ktsq\Bigr) & \to \delta(\ktsq)\,, \nn\\
 \theta \Big(t\dfrac{1-z}{z} - \ktsq\Big) \frac{1}{t} \cL_0\Bigl(\frac{\ktsq}{t}\Bigr) &\to \delta(\ktsq) \! \int_0^{\frac{1-z}{z}} \!\! \df r\, \cL_0(r) = \delta(\ktsq) \,\ln\!\Big(\frac{1\!-\!z}{z}\Big)\,,
  \nn\\
 \frac{1}{t} \cL_0 \Bigl(\frac{1\!-\!z}{z}-\frac{\ktsq}{t} \Bigr) & \to \delta(\ktsq) \! \int_0^{\frac{1-z}{z}} \!\! \df r\, \cL_0 \Bigl(\frac{1\!-\!z}{z}-r \Bigr) =  
\delta(\ktsq) \,\ln\!\Big(\frac{1\!-\!z}{z}\Big)
\end{align}
and the term in square brackets in \eq{exampleterms} vanishes. Hence we are free to replace $\cL_0[(1-z)/(1+z^2)] \to \theta(1-z) (1+z^2)/(1-z)$ in the splitting function $P^\zero_{qq}(z)$ multiplying this term without spoiling the behaviour of \eq{exampleterms} for $z\to1$.

Using the notation for the plus-distribution with the boundary condition at $r_{\rm max}=(1-z)/z$, as defined in appendix B of \mycite{Ligeti:2008ac}, we could also compactly express the terms in square brackets in \eq{exampleterms} as
\begin{align}
\frac{1}{t} \cL_0\Bigl(\frac{\ktsq}{t}\Bigr) -\ln\!\Big(\frac{1\!-\!z}{z}\Big)\, \delta(\ktsq) &= \frac{1}{t} \bigg[\frac{1}{r}\bigg]^{[r_{\rm max}]}_+ ,\nn\\
 \frac{1}{t} \cL_0 \Bigl(\frac{1\!-\!z}{z}-\frac{\ktsq}{t} \Bigr) - \ln\!\Big(\frac{1\!-\!z}{z}\Big)\, \delta\Bigl(t\frac{1\!-\!z}{z}-\ktsq\Bigr) &= \frac{1}{t} \bigg[\frac{1}{\frac{1-z}{z}-r} \bigg]^{[r_{\rm max}]}_+ ,
\end{align}
where $r\equiv \ktsq/t$.
For the sake of simplicity we however refrain from introducing another type of plus distribution and only use the one defined in \eq{plusdefmaintxt}, which has the boundary condition at 1, for the presentation of our results in the next section.

\section{Results}
\label{sec:results}

Here we present the results for the two-loop coefficient functions $J^\two_{ij}(t,z,\ktsq)$ in \eq{I2master} for $i = q$ and $i = \bar q$. Exploiting QCD charge conjugation invariance and in analogy to the functions $I^\two_{ij}(z)$ computed in \mycite{Gaunt:2014xga} we write 
\begin{align}
J_{\bar q_i \bar q_j}^\two = J_{q_i q_j}^\two
&= C_F\, \theta(z) \bigl[ \delta_{ij} J_{qqV}^\two + J_{qqS}^\two \bigr]
\,, \nn \\
J_{\bar q_i q_j}^\two = J_{q_i \bar q_j}^\two
&= C_F\, \theta(z) \bigl[ \delta_{ij} J_{q\bar qV}^\two + J_{qqS}^\two \bigr]
\,, \nn \\
J_{\bar q_i g}^\two = J_{q_i g}^\two
&= T_F\, \theta(z) \, J_{qg}^\two
\label{eq:Jij2}
\,,\end{align}
where $q_i$ ($\bar q_i$) denotes the (anti)quark of flavor $i$. With $r\equiv \ktsq/t$ we obtain
\begin{align} \label{eq:JqqV_two}
J_{qqV}^\two
&= \beta_0 \Biggl\{
    \bigg[ \Big(\frac{\pi^2}{6}  - \frac{14}{9}\Big) \delta(1\!-\!z)+\frac{5}{3}
    \Big(\cL_0(1\!-\!z) -\frac{1}{1\!-\!z} \Big)-\cL_1(1\!-\!z)+\frac{\ln (1\!-\!z)}{1\!-\!z}\bigg] \delta(\ktsq) \nn\\&\quad
    +\frac{5}{6}\frac{1\!+\!z^2}{1\!-\!z}\delta\Big(t \frac{1\!-\!z}{z}-\ktsq\Big)
    -\frac{1}{2}\frac{1\!+\!z^2}{1\!-\!z} \frac{1}{t}\cL_0\Big(\frac{1\!-\!z}{z}-r\Big) - \frac{r z^2+r+z^2+z}{2\, t\,(1\!-\!z) (r\!+\!1)^2}
  \Biggr\} \nn \\[1 ex] & 
 + C_A \Biggl\{
 \bigg[ \Big(7 \zeta (3)- \frac{16}{9} \Big) \delta(1\!-\!z)-\frac{\pi^2\!-4}{3} \Big( \cL_0(1\!-\!z)-\frac{1}{1\!-\!z} \Big) \bigg]\delta (\ktsq)  \nn\\&\quad
+ \bigg[\frac{1\!+\!z^2}{1\!-\!z} \Big( 2 \Li_2(z) - \frac{\pi ^2}{3} -\ln^2(1\!-\!z) -  \ln^2 z +4 \ln(1\!-\!z) \ln z \Big) + \frac{5 z^2 - 3 z + 2}{3(1\!-\!z)} \bigg] \nn\\&\qquad
 \times \delta\Big(t \frac{1\!-\!z}{z}-\ktsq\Big) 
+\frac{1\!+\!z^2}{1\!-\!z}\bigg[ 2  \ln\!\Big(\frac{1\!-\!z}{z}\Big) \frac{1}{t} \cL_0\Big(\frac{1\!-\!z}{z}-r\Big) -2\frac{1}{t} \cL_1\Big(\frac{1\!-\!z}{z}-r\Big) \bigg] \nn \\& \quad
+\frac{1\!+\!z^2}{1\!-\!z} \frac{1}{t} \bigg[\frac{ z \ln r}{1\!-\!r z\!-\!z} -\frac{\ln[1\!-\!(r\!+\!1) z]}{r\!+\!1} -\frac{(2 r z+2 z-1) \ln(1\!-\!z)}{(r\!+\!1) (1\!-\!r z\!-\!z)} \bigg] \nn\\& \quad
+\frac{1}{1\!-\!z} \frac{1}{t} \bigg[z -\frac{(r z^2+r+z^2+z) \ln z}{(r\!+\!1)^2} 
+\frac{(r z^2-2 r z-r+z^2-4 z+1) \ln(r\!+\!1)}{(r\!+\!1)^2 (1\!-\!r z\!-\!z)} \bigg]   
  \Biggr\}
   \nn \\[1 ex] & 
   + C_F \Biggl\{
 \bigg[8 \zeta(3)\delta(1\!-\!z) + 6 \cL_2(1\!-\!z)-\frac{5\pi ^2}{3} \Big( \cL_0(1\!-\!z) -\frac{1}{1\!-\!z} \Big) -\frac{6 \ln^2(1\!-\!z)}{1\!-\!z} \bigg] \delta(\ktsq) \nn\\& \quad
+\bigg[\frac{1\!+\!z^2}{1\!-\!z}\Big(\frac{\pi^2}{6}-2 \Li_2(z) -\ln^2 z +\ln^2(1\!-\!z) 
-\!4 \ln(1\!-\!z) \ln z \Big) +z \bigg]\delta\Big(t \frac{1\!-\!z}{z}-\ktsq\Big) \nn\\& \quad
-\frac{1\!+\!z^2}{1\!-\!z} \bigg[ 4 \ln\!\Big(\frac{1\!-\!z}{z}\Big)\frac{1}{t} \cL_0\Big(\frac{1\!-\!z}{z}-r\Big)
- 8 \frac{1}{t} \cL_1\Big(\frac{1\!-\!z}{z}-r\Big) -4 \frac{1}{t} \cL_1(r)  \bigg] \nn\\& \quad
+2(1\!-\!z) \frac{1}{t} \cL_0(r)
+\frac{r-r^2 z^2-2 r z^2+2 r z-z^2+5z-2}{t \,(1\!-\!z)(r\!+\!1)^2} - \frac{4(1\!+\!z^2) \ln(1\!-\!z)}{t \,(1\!-\!z)\, r} \nn\\& \quad
+\frac{\ln[1\!-\!(r\!+\!1) z]}{t \,(1\!-\!z)\,r\, (r\!+\!1)^2}\Big[4 (1\!+\!z^2)-r^3 (1\!-\!z) z+r^2(9 z^2\!-\!2 z\!+\!5) +4 r (3 z^2\!+\!2) \Big]  \nn\\& \quad
+\frac{\ln z \big[4 z (z^2\!-\!z\!+\!1)-r^3 (1\!-\!z) z^2+2 r^2 z (3 z^2\!-\!2 z\!+\!1)+ r (9 z^3\!-\!7 z^2\!+\!5 z\!+\!1)\big] }{t \,(1\!-\!z)(r\!+\!1)^2 (1\!-\!r z\!-\!z)}  \nn\\& \quad
-\frac{2 [r (z^2\!-\!2 z\!-\!1)+z^2-4 z+1] \ln(r\!+\!1)}{t \,(1\!-\!z)(r\!+\!1)^2 (1\!-\!r z\!-\!z)}
-\frac{2 (1\!+\!z^2) \ln r}{t \,(1\!-\!z)(r\!+\!1)}
 \Biggr\},
\end{align}
\begin{align} \label{eq:JqqbarV_two}
J_{q\bar qV}^\two
&= 
\frac{(2 C_F \!-\! C_A)}{t\,(1\!+\!z)(r\!+\!1)^2} \bigg\{\big[1+2 z-z^2 -r (1\!+\!z^2)\big] \ln(r\!+\!1)+\big[(1\!-\!z)z-r (1\!+\!z^2) \big] \ln z \nn\\& \quad
+(r\!+\!1) (1\!+\!z^2) \ln(1\!-\!r z)
\bigg\}
\,,\end{align}
\begin{align} \label{eq:JqqS_two}
J_{qqS}^\two
&= \frac{T_F}{t\,z\,(r\!+\!1)^4} \bigg\{
   r (8-3 z-7 z^2)-2 (1\!-\!z)^2 -r^2 (2+6 z+9 z^2)+ r^3 (z\!-\!5 z^2) - r^4 z^2 \nn\\& \quad
+z\, (r\!+\!1)^2 (z+2 r - z r^2) \ln z +4 \big[1-z+z^2+2 r z^2+r^2 (1+z+z^2)\big] \ln(r\!+\!1)   \nn\\& \quad
+(r^2\!+\!1) \big[2-2(r\!+\!1)z +z^2 (r\!+\!1)^2 \big] \ln[1\!-\!(r\!+\!1) z]
\bigg\}
\,,\end{align}
\begin{align} \label{eq:Jqg_two}
J_{qg}^\two
&= C_F \Biggl\{
\bigg[(2 z^2\!-\!2 z\!+\!1)  \Big( 2 \Li_2(z) - \ln^2(1\!-\!z)  + 4 \ln z \ln (1\!-\!z) -2 \ln^2 z - \frac{\pi^2}{6} \Big) + 7 z^2 - 8 z \nn\\& \qquad
 + \frac{7}{2} \bigg] \delta\Big(t \frac{1\!-\!z}{z}-\ktsq\Big) 
+(2 z^2\!-\!2 z\!+\!1) \bigg[
- \frac{3}{2} \frac{1}{t}\cL_0\Big(\frac{1\!-\!z}{z}-r\Big)
+2  \frac{1}{t} \cL_1\Big(\frac{1\!-\!z}{z}-r\Big) \nn\\& \qquad
+ 4  \frac{1}{t}  \cL_1(r) \bigg] +4 (1\!-\!z) z \frac{1}{t}  \cL_0(r) 
+\frac{\ln z}{t\,(r\!+\!1)^2 (1\!-\!r z\!-\!z)}\Big[2 r^4 z^3 - 2 r^3 (2-5 z) z^2 \nn\\& \qquad
 +r^2 z (5-20 z+26 z^2) -r (1-11 z+32 z^2-30 z^3) + 2 z (3 -8 z+6 z^2) \Big]  \nn\\& \quad
+\frac{\ln[1\!-\!(r\!+\!1) z]}{t\,r \, (r\!+\!1)^2} \Big[2 r^4 z^2-2 r^3 z (1\!-\!4 z) +2 r (10 z^2\!-\!8 z\!+\!3) + r^2 (18 z^2\!-\!10 z\!+\!3) \nn\\& \qquad
 +8 z^2 -8 z+4 \Big] 
+\frac{12 (1\!-\!2 z) z-16 r^3 z^2+14 r^2 (1\!-\!4 z) z+r (1\!+\!26 z\!-\!64 z^2)}{2\, t\, (r\!+\!1)^2} \nn\\& \quad
- \frac{2(2 z^2\!-\!2 z\!+\!1)}{t\,(r\!+\!1) (1\!-\!r z\!-\!z)} \bigg[ (1\!-\!2 r z\!-\!2 z) \ln r
+ (2\!-\!2 z\!+\!r\!-\!2 r z) \ln(1\!-\!z)\frac{1}{r } \bigg]
   \Biggr\}   \nn \\[1ex] &
   + C_A \Biggl\{
\bigg[(2 z^2\!-\!2 z\!+\!1) \Big( \ln^2(1\!-\!z) - 4 \ln z \ln(1\!-\!z) - 2 \Li_2(z) \Big) +z \bigg] \delta\Big(t \frac{1\!-\!z}{z}-\ktsq\Big) \nn\\& \qquad
+ (2 z^2\!-\!2 z\!+\!1) \bigg[- 2  \ln \Big( \frac{1\!-\!z}{z}\Big) \frac{1}{t} \cL_0\Big(\frac{1\!-\!z}{z}-r\Big) 
+4 \frac{1}{t}\cL_1\Big(\frac{1\!-\!z}{z}-r\Big) \bigg] \nn\\& \quad
 +\frac{1}{t\,z\,(r\!+\!1)^4}\Big[2 z^3\!-\!5 z^2\!+\!4 z\!-\!2 + 6 r^5 z^3  \!+\! r^4 z^2 (26 z\!-\!5) + r^3 z (44 z^2\!-\!18 z\!+\!1) \nn\\& \qquad
  + r^2 (36 z^3\!-\!26 z^2\!-\!6 z\!-\!2) + r (14 z^3\!-\!18 z^2\!-\!3 z\!+\!8) \Big] +  \frac{(2 z^2\!+\!2 z\!+\!1) \ln(1\!-\!r z)}{t\,(r\!+\!1)}  \nn\\& \quad
 +\frac{\ln(r\!+\!1)}{t\,z\,(r\!+\!1)^4 (1\!-\!r z\!-\!z)}  \Big[ 2 r^2 (6 z^4\!-\!18 z^3\!+\!3 z^2\!+\!2) + 4 r^3 z (2 z^3\!-\!2 z^2\!-\!z\!-\!1)    \nn\\& \qquad
  + 4 r z(2 z^3\!-\!10 z^2\!+\!7 z\!-\!1)+ z^2 r^4 (2 z^2\!+\!2 z\!+\!1) + 2 z^4 \!-\! 14 z^3 \!+\! 17 z^2 \!-\! 8 z \!+\! 4  \Big] \nn\\& \quad
 +\frac{ \ln[1\!-\!(r\!+\!1) z]}{t\,z\,(r\!+\!1)^4} \Big[\!-\!2 r^5 z^3 - 2 r^4 (3 z\!-\!1) z^2 - r^3 z(2 z^2\!+\!2 z\!+\!1) + r^2(10 z^3\!-\!10 z^2\!+\!z\!+\!2) \nn\\& \qquad
 +r z (12 z^2\!-\!6 z\!+\!1) + 4 z^3\!-\!z\!+\!2\Big]  
 + \frac{2 \ln z}{t\,(r\!+\!1)^2} \Big[r^3 z^2 + r^2 (z^2\!-\!z)-r z^2+r-z^2+z\Big] \nn\\& \quad
  -(2 z^2\!-\!2 z\!+\!1) \bigg[ \frac{(1\!-\! 2 r z \!-\!2z) \ln(1\!-\!z)}{t\,(r\!+\!1) (1\!-\!r z\!-\!z)}  +\frac{z  \ln r}{t\,(1\!-\!r z\!-\!z)} \bigg]
   \Biggr\}.
\end{align}
For simplicity we have suppressed the overall $\theta \big[ t (1\!-\!z)/z-\ktsq \big]$ factor already mentioned in section \ref{sec:match} (in all terms regular in the argument of this $\theta$-function). The combination of this function, the $\theta(z)$ in \eq{Jij2} and the support of the PDFs in \eq{dBi_OPE} enforces the kinematic constraints $(1-x)/x \ge r\equiv \ktsq/t \ge 0$ and $1 \ge x \ge 0$.
Also, we emphasize again, that due to the overall $\theta$-function the proper limits $z\to 1$ and $t\to0$ of the above results for the $J^\two_{ij}$ require the integration over $\ktsq$ (or equivalently $r$).

The expressions for the $J^\two_{ij}$ in eqs.~\eqref{eq:JqqV_two}-\eqref{eq:Jqg_two} as well as for the  $I^\two_{ij}$ in eqs.~(4.2)-(4.5) of \mycite{Gaunt:2014xga} are also available in electronic form upon request to the authors.

\section{Conclusions} 
\label{sec:conclusions}

In this paper we have presented the first NNLO calculation of a fully-unintegrated parton distribution, namely the (anti)quark dBF.
We have computed at two-loop order the matching coefficients $\cI_{qj}^\two(t,z,\ktsq,\mu)$ between the dBF $B_q(t,x,\ktsq,\mu)$ and the PDFs $f_j(x,\mu)$. We have checked our computation by using two different gauges, Feynman and axial light-cone gauge, and two different methods for taking the discontinuities of the operator diagrams that are required to obtain the partonic dBF matrix elements. 
Integration of $B_q(t,x,\ktsq,\mu)$ over the transverse momentum $\vec k_\perp$ yields the virtuality-dependent beam function $B_q(t,x,\mu)$.
Our results are an important ingredient to obtain the full NNLO singular contributions as well as the NNLL$'$ and N$^3$LL resummation for observables that probe both the virtuality and the transverse momentum of the colliding quarks.

\begin{acknowledgments}
We thank Frank Tackmann for comments on the manuscript.
Parts of the calculations in this paper and \mycite{Gaunt:2014xga} were performed using {\tt FORM}
 \cite{DBLP:journals/corr/abs-1203-6543}, {\tt HypExp} \cite{Huber:2005yg, Huber:2007dx} and {\tt FeynCalc} \cite{Mertig:1990an}.
The Feynman diagrams have been drawn using {\tt JaxoDraw}~\cite{Binosi:2008ig}.
This work was supported by the DFG Emmy-Noether Grant No. TA 867/1-1.
\end{acknowledgments}

\appendix

\section{Tree-level and one-loop matching coefficients}
\label{app:oneloopIs}


We define the expansion of the beam function matching coefficient as follows:
\begin{align}
\cI_{ij}(t,z,\ktsq,\mu) \,=\, \frac1\pi\, \sum_{n=0}^{\infty} \left(\dfrac{\alpha_s}{4\pi}\right)^{\!n} \cI_{ij}^{(n)}(t,z,\ktsq,\mu)
\,.\end{align}
The tree-level matching coefficients are
\begin{equation}
\label{eq:Iijtree}
\cI_{ij}^\zero(t,z,\ktsq,\mu) = \delta_{ij}\,\delta(t)\,\delta(\ktsq)\,\delta(1-z)
\,.\end{equation}
The one-loop matching coefficients are~\cite{Jain:2011iu}
\begin{align} \label{eq:Iijoneloop}
\cI_{ij}^\one(t,z,\ktsq,\mu)
&= \frac{1}{\mu^2} \cL_1\Bigl(\frac{t}{\mu^2}\Bigr) \Gamma_0^i\, \delta_{ij}\delta(\ktsq)\delta(1-z)
  + \frac{1}{\mu^2} \cL_0\Bigl(\frac{t}{\mu^2}\Bigr)
  \Bigl[- \frac{\gamma_{B\,0}^i}{2}\,\delta_{ij}\delta(\ktsq)\delta(1-z) \nn \\ & \quad
+ 2\, \delta\Big(t \frac{1\!-\!z}{z}-\ktsq\Big) P_{ij}^\zero(z) \Bigr]
  + \delta(t)\,\delta(\ktsq)\, 2I_{ij}^\one(z)
\,.\end{align}
%
The $\mu$-independent one-loop constants are
\begin{align}
I_{q_iq_j}^\one(z) &= \delta_{ij}\, C_F\, \theta(z) I_{qq}(z)
\,,\nn\\
I_{q_ig}^\one(z) &= T_F\, \theta(z) I_{qg}(z)
\,,\nn\\
I_{gg}^\one(z) &= C_A\, \theta(z) I_{gg}(z)
\,,\nn\\
I_{gq_i}^\one(z) &= C_F\, \theta(z) I_{gq}(z)
\,,\end{align}
with the quark matching functions~\cite{Stewart:2010qs} given by\footnote{Note that here $I_{ij}(z) \equiv \cI_{ij}^{(1,\delta)}(z)$ in the notation of \mycites{Stewart:2010qs, Berger:2010xi}.}
\begin{align} \label{eq:Iqdel_results}
I_{qq}(z)
&= \cL_1(1 - z)(1 + z^2)
  -\frac{\pi^2}{6}\, \delta(1 - z)
  + \theta(1 - z)\Bigl(1 - z - \frac{1 + z^2}{1 - z}\ln z \Bigr)
\,, \nn \\
I_{qg}(z)
&= P_{qg}(z)\Bigl(\ln\frac{1-z}{z} - 1\Bigr) +  \theta(1-z)
\,.\end{align}

\section{Perturbative ingredients}
\label{app:pert}

\subsection{Splitting functions} 
\label{app:splittingfunc}

We define the expansion of PDF anomalous dimensions ($\gamma^f_{ij}=2 P_{ij}$) in the $\overline{\mathrm{MS}}$ as follows:
\begin{align}
\label{eq:Pijexp}
P_{ij}(z,\alpha_s) = \sum_{n=0}^{\infty} \left(\dfrac{\alpha_s}{2\pi}\right)^{n+1} P_{ij}^{(n)}(z)\,.
\end{align}
The one-loop terms read
\begin{align}
P_{q_i q_j}^\zero(z) &= C_F\, \theta(z)\, \delta_{ij} P_{qq}(z)
\,,\nn\\
P_{q_ig}^\zero(z) = P_{\bar q_ig}^\zero(z) &= T_F\, \theta(z) P_{qg}(z)
\,,\nn\\
P_{gg}^\zero(z) &= C_A\, \theta(z) P_{gg}(z) + \frac{\beta_0}{2}\,\delta(1-z)
\,,\nn\\
P_{gq_i}^\zero(z) = P_{g\bar q_i}^\zero(z) &= C_F\, \theta(z) P_{gq}(z)
\,,\end{align}
with the usual one-loop (LO) quark and gluon splitting functions
\begin{align} \label{eq:Pij}
P_{qq}(z)
&= \cL_0(1-z)(1+z^2) + \frac{3}{2}\,\delta(1-z)
\equiv \biggl[\theta(1-z)\,\frac{1+z^2}{1-z}\biggr]_+
\,,\nn\\
P_{qg}(z) &= \theta(1-z)\bigl[(1-z)^2+ z^2\bigr]
\,,\nn\\
P_{gg}(z)
&= 2 \cL_0(1-z) \frac{(1 - z + z^2)^2}{z}
\,,\nn\\
P_{gq}(z) &= \theta(1-z)\, \frac{1+(1-z)^2}{z}
\,.\end{align}

\subsection{Convolutions of one-loop functions} \label{sec:oneloopconvs}

The (Mellin) convolution of two functions in the light-cone minus component is defined as
\begin{align}
\label{eq:convz}
f(z) \convz g(z) = \int_z^1\! \frac{\df w}{w}\, f(w) g \Bigl(\frac{z}{w}\Bigr)
\,,\end{align}
%
The convolutions of the one-loop splitting functions required in \eq{I2master} are ($r \equiv \ktsq/t$):
\begin{align}
\label{eq:convPqqPqq}
\Big[&\delta\Bigl(t\frac{1\!-\!z}{z}-\ktsq\Bigr) P_{qq}(z) \Big]  \conv_z   P_{qq}(z) = 
2\frac{z^2+1}{1-z} \Big[\frac{1}{t} \cL_0(r) + \frac{1}{t}\cL_0\Big(\frac{1\!-\!z}{z}-r\Big) \Big]
\nn\\
&+\delta(\ktsq) \bigg[\Big(\frac{9}{4}-\frac{2 \pi ^2}{3} \Big) \delta(1-z) + 6 \cL_0(1-z) + 8 \cL_1(1-z) +\frac{3 z^2-16 \ln(1-z)-9}{2
   (1-z)} \bigg] \nn\\
&-\frac{(z^2+1) (4 \ln z + 3)}{2 (z-1)} \delta\Big(t \frac{1\!-\!z}{z}-\ktsq\Big) +\frac{(r+2) (z-1) (r z+z+1)}{t\,(1-z) (r+1)^2}\,, \\[1 ex]
\Big[&\delta\Bigl(t\frac{1\!-\!z}{z}-\ktsq\Bigr) P_{qq}(z) \Big]  \conv_z   P_{qg}(z) = 
2 P_{qg}(z) \frac{1}{t} \cL_0(r) + \dfrac{3}{2}P_{qg}(z) \delta(\ktsq) \nn\\
& +\frac{2 r^3 z^2+2 r^2 z (4 z-1)+r (10 z^2-2 z-1)+4 z^2-2}{t\,(r+1)^2}\,, \\[1 ex]
\Big[&\delta\Bigl(t\frac{1\!-\!z}{z}-\ktsq\Bigr) P_{qg}(z) \Big]  \conv_z   P_{gq}(z) =  
\frac{(r^2+1) \big[ (r+1)^2 z^2-2 (r+1) z+2 \big]}{t\,(r+1)^4 z}\,, \\[1 ex]
\Big[&\delta\Bigl(t\frac{1\!-\!z}{z}-\ktsq\Bigr) P_{qg}(z) \Big]  \conv_z   P_{gg}(z)  =  
2 P_{qg}(z) \Big[ \dfrac{1}{t} \cL_0\Big(\frac{1\!-\!z}{z}-r\Big) + \ln z \, \delta\Big(t \frac{1\!-\!z}{z}-\ktsq\Big) \Big] \nn\\
& -2\frac{r^5 z^3 + r^4 z^2(3 z\!-\!1) + r^3(2 z^3\!+\!z) - r^2 (2 z^3\!-\!2 z^2\!-\!z\!+\!1) + r(z\!-\!3 z^3)-z^3\!-\!z^2\!+\!z\!-\!1 }{t\,(r+1)^4 z}
\,.
\label{eq:convPqgPgg}
\end{align}
In \eqs{convPqqPqq}{convPqgPgg} we have again suppressed an overall $\theta \big[ t (1\!-\!z)/z-\ktsq \big]$ factor multiplying all terms regular in the limit $\ktsq\to t(1-z)/z$.

\section{Plus distributions}
\label{app:Useful}

We define the standard plus distributions as
\begin{align} \label{eq:plusdef}
\cL_n(x)
&= \biggl[ \frac{\theta(x) \ln^n x}{x}\biggr]_+
 = \lim_{\eps \to 0} \frac{\df}{\df x}\biggl[ \theta(x- \eps)\frac{\ln^{n+1} x}{n+1} \biggr]
\,.\end{align}
For the derivation of \eq{I2master} from \eq{IRGE} we need the derivatives
\begin{align}
\mu\frac{\df}{\df\mu}\,\frac{1}{\mu^2}\cL_n\Bigl(\frac{t}{\mu^2}\Bigr)
&= -2n\, \frac{1}{\mu^2}\cL_{n-1}\Bigl(\frac{t}{\mu^2}\Bigr)
\qquad (\forall\, n \ge 1)
\,, \nn \\ \label{eq:Lnderivs}
\mu\frac{\df}{\df\mu}\,\frac{1}{\mu^2}\cL_0\Bigl(\frac{t}{\mu^2}\Bigr)
&= -2\delta(t)
\,.\end{align}
%

\bibliographystyle{../jhep}
\bibliography{../beamfunc}

\end{document}